\documentclass[aip,jcp,preprint]{revtex4-2}

\usepackage{graphicx}

\usepackage{amsmath}
\usepackage{amssymb}
\usepackage{subcaption}
\usepackage{hyperref}
\usepackage{booktabs} 
\usepackage{multirow}

\begin{document}

\title{Thermodynamics and Kinetics of a Three-Arm Star Polymer Translocating through a Nanopore}

\author{Bhavesh R. Sarode}
\author{Harshwardhan H. Katkar}
\email[Author to whom correspondence should be addressed: ]{hkatkar@iitk.ac.in}
\affiliation{FB 457, Department of Chemical Engineering, Indian Institute of Technology Kanpur, Kanpur - 208016, Uttar Pradesh, India}


\begin{abstract}
In this work, voltage-driven translocation of uniformly charged long linear and three-arm star polymers through narrow nanopores is investigated. Langevin dynamics simulation is performed using a coarse-grained model of the polymer and a semi-implicit representation of the nanopore. The mean translocation time of a linear polymer is found to be inversely proportional to the applied voltage over a wide range of voltages. In contrast, the mean translocation time of a three-arm star polymer of the same molecular weight exhibits a pronounced deviation from this scaling relation below a threshold voltage. The threshold voltage is found to be nearly independent of the molecular weight of the polymer, but depends on the size of the nanopore and salt concentration. Metadynamics simulation is used to estimate the free-energy landscape for the translocation of the three-arm star polymer. Below the threshold voltage, the free energy exhibits a pronounced second barrier resulting from an entropic contribution and electrostatic interactions between segments of the trailing arm inside the nanopore. A Fokker-Planck model developed using the estimated free-energy accurately predicts the deviation from the scaling relation below the threshold voltage and shows a remarkable agreement with the Langevin dynamics simulation results using a voltage-independent fitting parameter. The agreement between the theory and the Langevin dynamics simulation results is seen for different nanopore radii, molecular weights of the polymer and salt concentrations studied. A simple extension of the free energy landscape is suggested to predict translocation kinetics for higher molecular weights of the polymer without performing additional computationally expensive simulations.

\end{abstract}
\maketitle 
\section{Introduction}
\label{sec:intro}

Driven translocation of charged macromolecules through nanopores is ubiquitous in nature. \cite{Muthukumar2011PolymerTranslocation} It also finds applications in single-molecule detection of DNA mutations, proteinopathy, characterizing synthetic polymers, etc. \cite{Kasianowicz1996,Bezrukov1994,Tripathi2021} Synthetic polymers of various topologies, including star polymers, are used in industrial applications. Translocation through nanopores can be used as a robust method for their molecular characterization. Kinetics of translocation of polymers through nanopores is typically reported in terms of the mean translocation time $\langle \tau \rangle$ or a dwell time, a key parameter measured in experiments.

Several experimental investigations into voltage-driven translocation of linear polymers have reported a scaling of $\langle \tau \rangle \sim 1/V$, where $V$ is the transmembrane voltage. \cite{Storm2005,Wanunu2008,Brun2008,Carson2014,Liu2012,Chen2024,Wong2010} 
Storm \textit{et al.} \cite{Storm2005} report this scaling for translocation of 6557 to 48500 base-pair long dsDNA through a 10 nm SiN nanopore. Wanunu et al. \cite{Wanunu2008} investigated the translocation of 150--3500 bp DNA through solid-state nanopores with diameters ranging from 2.7 to 5 nm and report this scaling. Brun \textit{et al.} \cite{Brun2008} studied the voltage-driven translocation of dextran sulfate sodium through an $\alpha$-hemolysin nanopore under varying applied transmembrane voltages. Carson \textit{et al.} investigated the translocation of 35--20,000 bp single-stranded DNA (ssDNA) through a 5 nm silicon nitride (SiN) nanopore. Liu et al. \cite{Liu2014} examined the translocation of 48.5 kbp $\lambda$-DNA through a SiN nanopore. Wong \textit{et al.} \cite{Wong2010} studied the translocation of sodium polystyrene sulfonate (NaPSS) through an $\alpha$-hemolysin nanopore. Despite the differences in polymer type, molecular weight, and nanopore geometry, all of these experimental studies report that the mean translocation time scales inversely with the applied transmembrane voltage over a wide range of polymer lengths in both solid-state and biological nanopores. Higher voltages applied across nanopores can result into a change in the degree of ionization of the polymer, resulting into deviation from this scaling behavior for linear polymers. \cite{Luo2006,Jeon2016,Wei2025,Colchero2026,Kowalczyk2012a} For example, a scaling of $\langle \tau \rangle \sim \exp\left(-V/V_0\right)$ has also been reported for small solid-state nanopores. \cite{Wanunu2008,Brun2008,Chen2010} 

The same inverse-voltage scaling of the mean translocation time with voltage has also been observed in simulation of linear polymers translocating through nanopores.\cite{Luo2006,Lehtola2008,Dubbeldam2007,Metzler2010} Using two-dimensional Monte Carlo simulations, Luo \textit{et al.} \cite{Luo2006} reported a slightly different power-law dependence of $\langle \tau \rangle \sim V^{-0.83}$ at low applied electric fields. Lehtola \textit{et al.} \cite{Lehtola2008} investigated polymer translocation using both Monte Carlo and Langevin dynamics simulations and observed the scaling $\langle \tau \rangle \sim V^{-1}$. Dubbeldam \textit{et al.} \cite{Dubbeldam2007} also reported the same inverse-voltage scaling from Monte Carlo simulations, while Metzler \textit{et al.} \cite{Metzler2010} confirmed this behavior using three-dimensional Langevin dynamics simulations of linear polymers.

While several experimental and simulation studies have reported this scaling for linear polymers, limited work has been done to investigate voltage-dependent kinetics of translocation of star polymers. Using three-dimensional coarse-grained Langevin dynamics simulations, Katkar and Muthukumar \cite{Katkar2018a} investigated the voltage-driven translocation of charged star polymers through nanopores and demonstrated that the mean translocation time depends strongly on both polymer functionality and pore geometry. They reported a non-monotonic dependence of the mean translocation time on polymer functionality over the range $f=2-10$. Tilahun and Tatek \cite{Tilahun2023} extended these studies to uncharged self-avoiding star polymers with functionalities ranging from $f=2$ to $12$. They also reported a non-monotonic dependence of the mean translocation time on functionality of the star polymer and established the scaling relation $\langle \tau \rangle \sim N^{\alpha}V^{-1}$, where the exponent $\alpha$ depends on the polymer functionality. In an earlier study, \cite{Tilahun2021} the same authors had investigated the escape of star polymers through a cylindrical cavity and reported a similar non-monotonic behavior with functionality and the inverse-voltage scaling, $\langle \tau \rangle \sim V^{-1}$. Nagarajan and Chen \cite{Nagarajan2019b} investigated the flow-driven translocation of star polymers through nanopores and  reported a non-monotonic dependence of the mean translocation time on functionality. In another study, \cite{Nagarajan2019a} dissipative particle dynamics simulations were used to investigate the voltage-driven translocation of charged star polymers through cylindrical nanopores and confirmed the same inverse-voltage scaling. They further reported that the likely translocation pathway corresponds to a smaller number of arms leading the translocation process, with the ratio of forward to backward arms being approximately 0.25, in agreement with the experimental observations of Chen and Muthukumar \cite{Chen2024}. The non-monotonic dependence of the mean translocation time on polymer functionality was also reported using Langevin dynamics simulation of star polymers translocating into a spheroidal cavity.  \cite{Nagarajan2020} Liu \textit{et al.} \cite{Liu2014} demonstrated that the translocation dynamics of star polymers are strongly influenced by their initial conformation. For three-arm star polymers, the two-arm-forward configuration exhibits a shorter mean translocation time than the one-arm-forward configuration, indicating different values of the scaling exponent $\alpha$. Nevertheless, both simulations and experiments have shown that, when approaching a narrow nanopore, star polymers preferentially adopt conformations with fewer leading arms, making the one-arm-forward configuration the dominant translocation pathway. \cite{Katkar2018a,Chen2024}

1-dimensional Fokker-Planck theory has been used to describe the translocation of the linear polymers through the narrow confinement.\cite{Sung1996,Muthukumar1999,Muthukumar2003, Katkar2014, Katkar2018b} The theory relies on two inputs, \textit{viz.} the free energy landscape of the translocation process and the diffusivity parameter. Muthukumar \cite{Muthukumar2003} suggested an analytical model of the free energy of a linear polymer translocating through a finite-length nanopore, highlighting the presence of an entropic barrier. The diffusivity has been suggested to be constant, dependent on the molecular weight of the polymer, or reaction-coordinate-dependent. \cite{Sung1996,Muthukumar1999,Polson2014,Sarode2025}  For flow-driven translocation of a three-arm star polymer through a hole, Liu \textit{et al.} simplified the analytical free energy proposed by Muthukumar \cite{Muthukumar1999} by ignoring conformational entropy of the translocating polymer any intra-polymer interactions, and predicted an inverse scaling relationship between translocation time and the driving chemical potential difference. However, no theory has been developed to predict the voltage-dependent translocation of a star polymer through a finite-length nanopore.

Here, we use a combination of Langevin dynamics simulations, metadynamics and the Fokker-Planck theory to investigate the thermodynamics and kinetics of translocation of three-arm star polymer through a nanopore. Simulation details and theory are described in the following section. Section \ref{sec:results} begins with a discussion about the $\langle\tau\rangle\propto V^{-1}$ scaling for linear polymers and deviation from this scaling for three-arm star polymers, followed by insights obtained from the underlying free energy landscape. Effect of molecular weight of the polymer, size of the nanopore and salt concentration on the free energy are also discussed. A comparison of the translocation time from Langevin dynamics simulation and the Fokker-Planck theory is presented. A simple extension of the free energy is suggested at the end.

\section{Methods}
\label{sec:methods}
Reduced units are used for all quantities by choosing a lengthscale of $3$ \AA, a mass scale of $130$ g/mol, and an energy scale of $k_BT$ with $T=300$ K. Here $k_B$ is the Boltzmann constant. The resulting timescale $t_s = \sigma\sqrt{\frac{m}{\epsilon}} = 2.17$ ps. All quantities reported in the remainder of this manuscript are in reduced units, unless explicitly stated otherwise.

\subsection{Simulation}
\label{sec:methods:simulation}

\begin{figure}
    \centering
    \includegraphics[width=\linewidth]{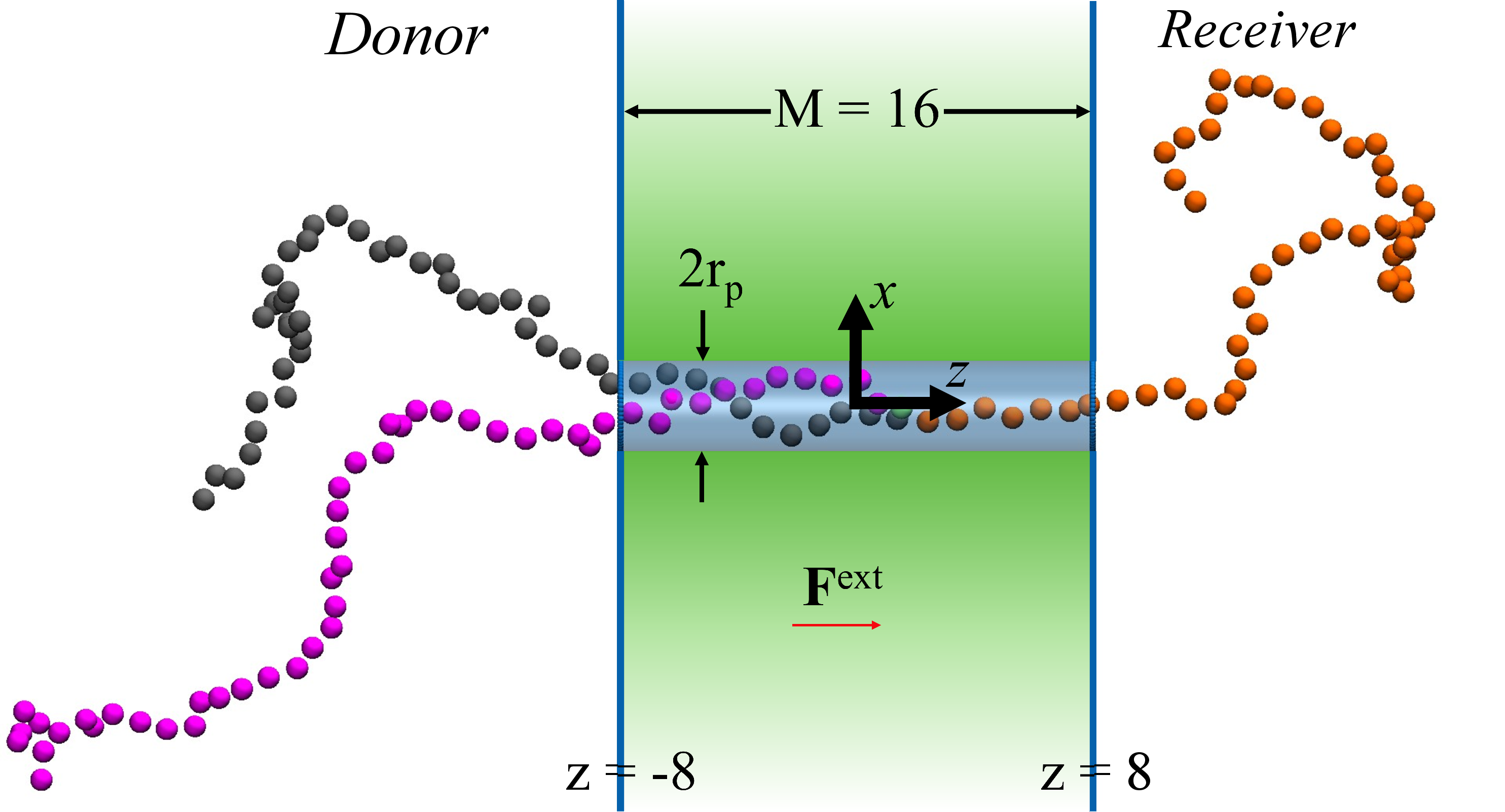}
    \caption{Schematic of the Langevin dynamics simulation setup showing a charged star polymer of length $N=151$ and functionality $f(=3)$ with one leading arm (orange beads) and $f-1(=2)$ trailing arms (grey and magenta beads) translocating through the nanopore of radius $r_p=1.4$ and length $M=16$ from donor to receiver compartment in the presence of externally applied force $\textbf{F}^\text{ext}$. The membrane region is between $z=-8$ and $z=8$.}
    \label{fig:systemdescription}
\end{figure}

A homogeneous star polymer with a total of either $N=151$ or $N=241$ uniform beads is considered in this study. The mean radii of gyration of these polymers are 16.3 and 22.0, respectively. A 3-dimensional periodic simulation box with $x$ and $y$ dimensions greater than 3 times the radius of gyration of the polymer is used. The box length in $z$ dimension is 200 (for $N=151$) or 240 units (for $N=241$). The box is partitioned into donor and receiver compartments by a membrane of thickness $M=16$, and a cylindrical nanopore of radius $r_p$ is embedded at the center of the box, as shown in Figure \ref{fig:systemdescription}.

The polymer consists of $f$ linear arms, each made of $\ell= (N-1)/f$ connected beads. $f=2$ results into a linear polymer, while $f=3$ results into a three-arm star polymer. These arms are connected to a central bead, which serves as a branch point. The branch point is electrically neutral, whereas the remaining beads possess a negative charge of -13.63, which corresponds to an elementary charge. Each bead has a unit mass.

\subsubsection{Langevin Dynamics Simulation}
\label{sec:methods:LD}

The position vector $\mathbf{r_i}$ of the $i^\text{th}$ polymer bead undergoes dynamics according to the Langevin equations of motion, given by

\begin{align}
    m_i \frac{d^2 \mathbf{r}_{i}}{dt^2} =  - \mathbf{\nabla} (U_{\mathrm{LJ}} + U_b + U_{\mathrm{DH}} ) -\zeta \frac{d \mathbf{r}_{i}}{dt} + \mathbf{F}_{i}^{\text{R}} + \mathbf{F}^{\text{ext}},
    \label{eq:LD1}
\end{align}

where $m_i$ is the mass of the polymer bead and $t$ is time. All interactions and cutoffs are similar to those in our previous work.\cite{Sarode2025} Briefly, conserved forces on the bead are modeled using three potentials, i.e., pairwise Lennard-Jones interaction $U_\text{LJ} = 4\epsilon_\text{LJ}\left[\left(\frac{\sigma_\text{LJ}}{r_{ij}}\right)^{12}-\left(\frac{\sigma_\text{LJ}}{r_{ij}}\right)^{6}\right]$, Debye-Hückel electrostatic interactions $U_\text{DH} = \frac{q_iq_j}{80r_{ij}}\exp\left[-\kappa r_{ij}\right]$, and harmonic bond potential $U_b = 15480\left(r_{ij} - 1\right)^2$. Here, $r_{ij}$ is the distance between two beads with charges $q_i, q_j$, and the potential $U_b$ is only used for connected beads. $\kappa$ is the inverse Debye length. For a 0.1 M solution of a monovalent salt in water, $\kappa = 0.3080$. The fluctuation-dissipation theorem governs the relationship between the drag force, characterized by a drag coefficient $\zeta = 1$, and the random force $\mathbf{F}^R$ acting on the monomer due to the implicit solvent. The membrane walls and the cylindrical nanopore are implicitly incorporated using the \textit{fix region} command and LJ walls in LAMMPS, \cite{PLIMPTON19951} with two additional rings of static neutral beads at the two ends of the nanopore, resulting into a semi-implicit nanopore. $\sigma_\text{LJ}=1$ for polymer beads, while for the LJ walls and the static beads of the nanopore, $\sigma_\text{LJ} = 0.25$. The distance between neighboring static beads in each ring is kept below 0.25. The interaction $U_\text{LJ}$ is truncated at $2^{1/6}\sigma_\text{LJ}$, and $\epsilon_\text{LJ} = 1$. A voltage difference of $V$ applied across the nanopore is assumed to result in a uniform electric field of magnitude $E = V/M$ in the $z$-direction inside the nanopore. The resulting external force on a charged polymer bead $i$ located inside the nanopore is given by $\mathbf{F}^{\text{ext}} = q_i \mathbf{E}$.

\subsubsection{Simulation Protocol}
\label{sec:methods:LD:Simulation}
The system described above was simulated in LAMMPS using the following protocol. The origin was placed at the center of the periodic box. The membrane extended from $z=-8$ to $z=8$ with its center at the origin, with its surface normal in the $z$ direction. The cylindrical nanopore was embedded in the membrane with its axis aligned along the $z$ direction and passing through the origin. The initial conformation of the polymer was such that first bead of one (leading) arm was located just inside the nanopore at $(0, 0, -7)$, while positions of the remaining beads were generated randomly, ensuring that they were located in the donor compartment and bonded beads were kept at a distance of 1 from each other.

To generate an ensemble of equilibrium chain conformations, a Lennard-Jones wall of $\sigma=0.25$ at $(0,0,-8)$ was used instead of the semi-implicit nanopore. The first polymer bead was fixed at its initial location while the remaining polymer beads underwent Langevin dynamics in the absence of an external force. The interactions between the first two polymer beads and the Lennard-Jones wall were turned off during equilibration. Initial velocities based on $T=1$ were assigned to all remaining polymer beads. The Langevin equations of motion (\ref{eq:LD1}) were solved numerically using the Velocity-Verlet algorithm with a time step $dt = 0.005$. Total simulation time to generate equilibrium chain conformations was ensured to be longer than the Rouse relaxation time of a linear polymer of length $N$, given as
\begin{align*}
\tau_R = \frac{\zeta  \sigma^2 N^2}{3 \pi^2 k_B T} \approx 3.3774 \times 10^{-2} N^2
\end{align*}

2000 statistically independent equilibrium chain conformations were generated for each polymer length $N$. Following this, the Lennard-Jones wall used during the equilibration simulation was replaced with the semi-implicit nanopore. All beads of the polymer, including the first bead, underwent Langevin dynamics in the presence of a uniform electric field. A simulation run was considered to be successful if the entire polymer chain crossed the nanopore and exited into the receiver compartment. 

Translocation time $\tau_\text{LD}$ was exclusively determined for successful simulation runs and is defined as the duration between the last time the polymer chain enters the pore and the first time it exits the pore. The probability of success $P(success)$ is defined as the fraction of the 2000 simulation runs that were successful.

\subsection{Fokker-Planck Formalism}\label{sec:methods:fokker_planck}

The process of driven translocation of a star polymer is modeled using a one-dimensional Fokker-Planck formalism, previously used for linear polymers \cite{Sung1996,Muthukumar1999,Muthukumar2003,Katkar2014, Polson2014,Sarode2025}. A single reaction coordinate $x$ is used to describe different ``states'' of the system as $N$ beads of the polymer chain translocate across the nanopore of length $M$. 

\begin{figure}
    \centering
    \includegraphics[width=\linewidth]{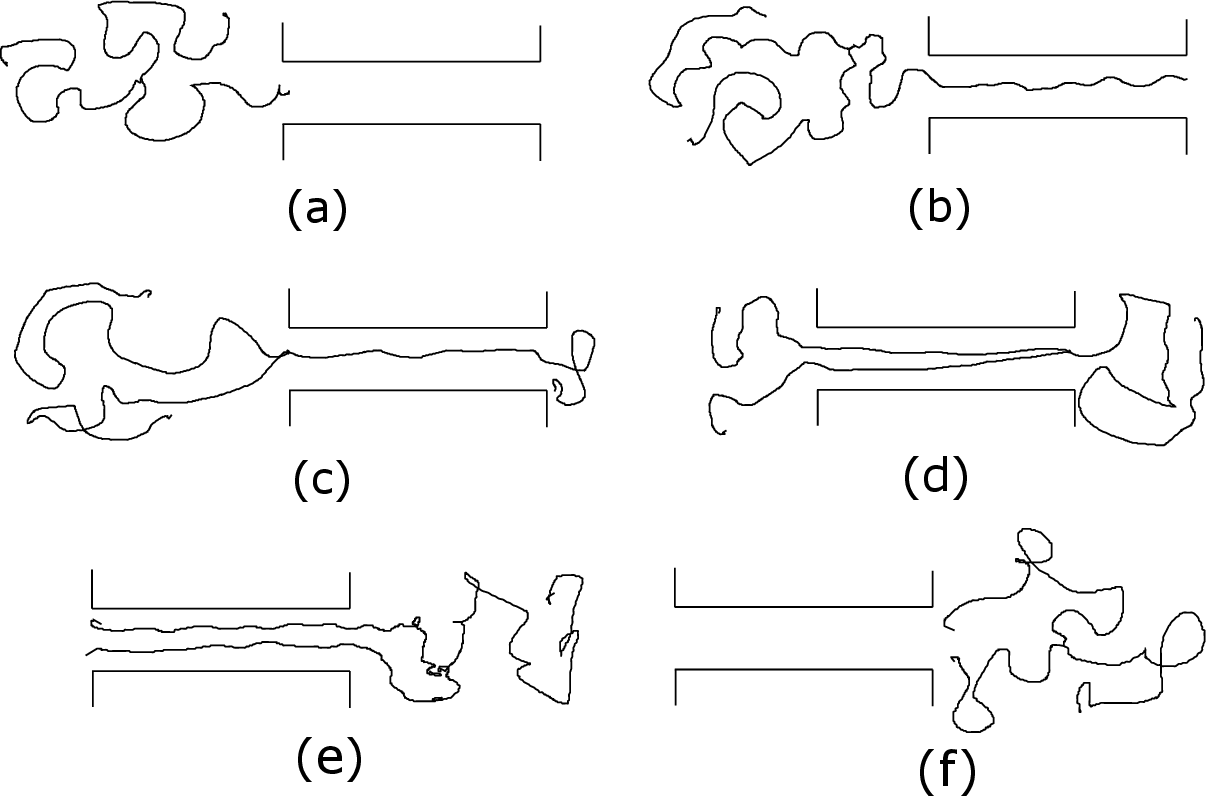}
    \caption{Schematic illustration of the translocation pathway of a three-arm star polymer through a nanopore, divided into five distinct stages: (a-b) pore-filling stage, (b-c) branch-point adsorption stage, (c-d) branch-point transfer stage, (d-e) polymer transfer stage, and (e-f) pore-emptying stage.}
    \label{fig:schematic_trans_progress_curve}
\end{figure}

The process is divided into five stages: pore filling (a-b), leading-arm threading (b-c), branch-point transfer (c-d), trailing-arm threading (d-e), and pore emptying (e-f), as shown in the Figure \ref{fig:schematic_trans_progress_curve}. The reaction coordinate is equal to the count of polymer beads crossing the entrance of the nanopore until the beginning of the pore emptying stage. Here, the reaction coordinate is equal to $N$ plus the length of the nanopore emptied by each trailing arm of the polymer. Thus, the pore filling stage spans $x\in[0,M]$, leading-arm threading stage spans $x\in[M,\ell]$, branch-point transfer stage spans $x\in[l,l+(f-1)M]$, trailing-arm threading stage spans $x\in[l+(f-1)M,N]$ and the pore emptying stage spans $x\in[N,N+(f-1)M]$. Thus, the reaction coordinate spans $x\in[0,N+(f-1)M]$.

The probability $W(x,t)$ of observing the system in state $x$ at time $t$ evolves according to the Fokker-Planck equation,
\begin{align*}
\frac{\partial W(x,t)}{\partial t} &= \frac{\partial J(x,t)}{\partial x}, & \text{with} \nonumber\\
J(x,t) &= \frac{k_\text{FP} }{k_B T}  \frac{\partial F'(x)}{\partial x} W(x,t)  +  k_\text{FP} \frac{\partial W(x,t)}{\partial x},
\end{align*}
where $F'$ is the dimensional free energy.
Rearranging,
\begin{align}
\frac{\partial W(x,t)}{\partial \tilde{t}} = \frac{\partial W(x,t)}{k_\text{FP}\partial t} &= \frac{\partial}{\partial x} \left[ \frac{\partial F(x)}{\partial x} W(x,t)  +  \frac{\partial W(x,t)}{\partial x}\right], \nonumber\\
\label{eq:non_dimen_FPE}
\end{align}
where $\tilde{t} = k_\text{FP}t$ is the rescaled time and $F = \frac{F'}{k_BT}$ is the dimensionless free energy. The system's ``diffusivity" $k_\text{FP}$ (s$^{-1}$) is assumed to be uniform along the reaction coordinate. The net flux $J(x,t)$ consists of drift arising from the local gradient of the free energy and diffusion fluxes.

\subsubsection{Metadynamics simulation}
\label{sec:methods:FP:metad}

LAMMPS \cite{PLIMPTON19951} patched with the PLUMED version 2.7.6 \cite{bonomi2009plumed} was used to estimate the free energy landscape $F_{meta}$ for translocation of a three-arm star polymer through a nanopore in the absence of an electric field using untempered metadynamics.\cite{Laio2002} The reaction coordinate $x$ was made differentiable by approximating it using a sum of error functions, as
\begin{align*}
 x &= \sum_{i=1}^{N-1} \frac{1 + \text{erf}\left(4(z_i + 8) - 2\right)}{2} + \sum_{j=2}^{f}\sum_{z'=-8}^8 \frac{1 + \text{erf}\left(4(z_{f,j} - z') - 2\right)}{2}, &\text{where}\\
 \end{align*}
Here, $z_i$ is the $z-$coordinate of $i^{th}$ bead, while $z_{f,j}$ is the $z-$coordinate of the last bead of $j^{th}$ trailing arm. The first summation gives a count of all the polymer beads that have crossed the nanopore entrance, while the second gives the total length of the empty part of the nanopore.

A Gaussian bias of height $h=0.005$ and width $\sigma_\text{meta} = 0.7$ was added every 2.5 time units to enhance sampling along the reaction coordinate $x$. We introduced additional constraint potentials to limit sampling of the reaction coordinate in the range of our interest and to prevent bending of arms inside the nanopore. The first four are given as $1000(x-3)^2$ for $x<3$, $1000(x-(N+2M-6))^2$ for $x>N+2M-6$, $1000(z_1+5)^2$ for $z_1<-5$ and $1000(z_{f,j}-5)^2$ for $z_{f,j}>5$. An additional constraint potential $0.016(n_2-n_3)^2$, where $n_j$ is the total number of beads of the trailing arm $j$ in the receiver compartment, as described in section SI1 of the supplementary document. Statistically independent metadynamics simulation runs were initiated using 5 of the 2000 equilibrium polymer conformations and semi-implicit nanopore in absence of any electric field. Details of the convergence criteria used in our work are discussed in section SI1 of the supplementary document. The average of converged free energy estimates from the five metadynamics simulation runs is reported as $F_\text{meta}$.

The voltage contribution to free energy $F^\text{volt}(x)$ for a star polymer translocating through a narrow nanopore, given that $\ell \geq M$, is

\begin{align}
    F_\text{volt} &=
    \begin{cases}
        qE \, \frac{x^2}{2} & \text{pore filling} \\
        qE\frac{M^2}{2} + \mu(x - M) & \text{leading-arm threading}\\
        qE\left[\frac{M^2}{2} + \frac{1}{2}(f-2)\left(\frac{x - \ell}{f - 1}\right)^2\right] \\
         \quad + \mu \left(x - \frac{(f-2)(x-\ell)}{f-1} + M\right) & \text{branch-point transfer} \\
        qE(f-1)\frac{M^2}{2} + \mu\left(x - (f-1)M\right) & \text{trailing-arm threading}\\
        qE(f-1)\left[\frac{M^2}{2} - \frac{(N+M - x)^2}{2}\right] \\
         \quad + \mu \big(x - (f-1)M\big) & \text{pore emptying}
    \end{cases}
    \label{eq:Fvolt}
\end{align}

Here, $E=V/M$ is the magnitude of the electric field and $\mu=qV$ is the electrochemical energy per segment in the receiver compartment relative to the donor compartment. In each stage, the first term on the right-hand side results from the integral over $z-$coordinates of all beads inside the nanopore, while the second term is related to the total number of beads in the receiver compartment. 

The resulting free energy $F = F_\text{meta}+F_\text{volt}$ is used in equation \ref{eq:non_dimen_FPE} with absorbing boundary conditions, \textit{i.e.} $W(x=0,t)=W(x=N+(f-1)M,t)=0$. The initial condition is approximated as $W(x,t=0) = \frac{1}{\sqrt{2\pi}\sigma}\exp{-(x-x_0)/2\sigma^2}$ with $\sigma=0.3$ and with $x_0=2$, which is consistent with the initial conformation of the polymer chain used in the Langevin dynamics simulation as described in Section \ref{sec:methods:LD:Simulation}. Equation \ref{eq:non_dimen_FPE} is solved for $W(x,t)$ using the method of lines approach implemented in the MATLAB \cite{MATLAB} function \textit{ode15s}, with a numerical upwinding scheme.  Section SI2 of the supplementary document provides details of the numerical scheme used.

The distribution $g(\tilde{\tau}_\text{FP})$, mean $\langle \tilde{\tau}_\text{FP} \rangle$ and standard deviation $\tilde{\sigma}_{\text{FP}}$ of the rescaled exit time $\tilde{\tau}_\text{FP} = k_\text{FP}\tau_\text{FP}$ are related to the probability flux at the boundary \( x=N+(f-1)M \) as \cite{Gardiner, Muthukumar2011PolymerTranslocation}

\begin{align}
g(\tilde{\tau}_\text{FP}) &= \frac{d}{d\tilde{\tau}_\text{FP}} \left(1- \frac{\int_{\tau_\text{FP}}^{\infty} J(N+(f-1)M,t) \, dt}{\int_{0}^{\infty} J(N+(f-1)M,t) \, dt} \right) \label{eq:dist_fpt} \\
\langle \tilde{\tau}_\text{FP} \rangle &= k_\text{FP}\langle \tau_{\text{FP}} \rangle = \int_{0}^{\infty} \tilde{t} \, g(\tilde{t})  d\tilde{t} \label{eq:fpt}\\
\tilde{\sigma}_{\text{FP}} &= \sqrt{ \left[\int_{0}^{\infty} \tilde{t}^2 \, g(\tilde{t}) d\tilde{t} \right] -  \langle \tilde{\tau}_\text{FP} \rangle^2 }.
\label{eq:sd_fpt}
\end{align}

\section{Results and Discussion}\label{sec:results}

\subsection{Translocation Kinetics: Langevin Dynamic Simulation}
\label{sec:translocatio_kinetics_LD}
\begin{figure}
    \centering
    \begin{subfigure}[b]{0.48\linewidth}
        \centering
        \includegraphics[width=\linewidth]{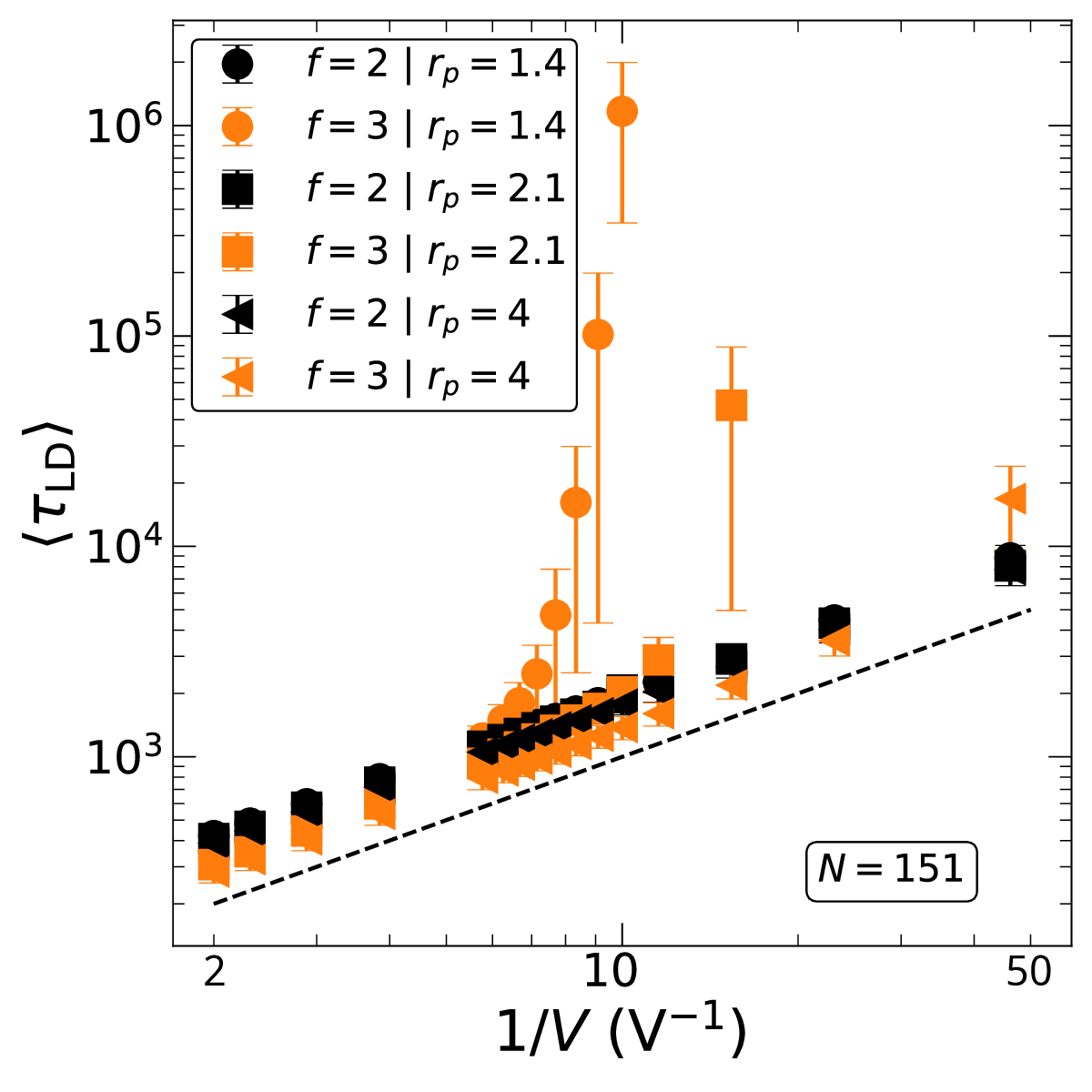}
        \caption{ }
        \label{fig:tau_LD_N151}
    \end{subfigure}
    \hfill
    \begin{subfigure}[b]{0.48\linewidth}
        \centering
        \includegraphics[width=\linewidth]{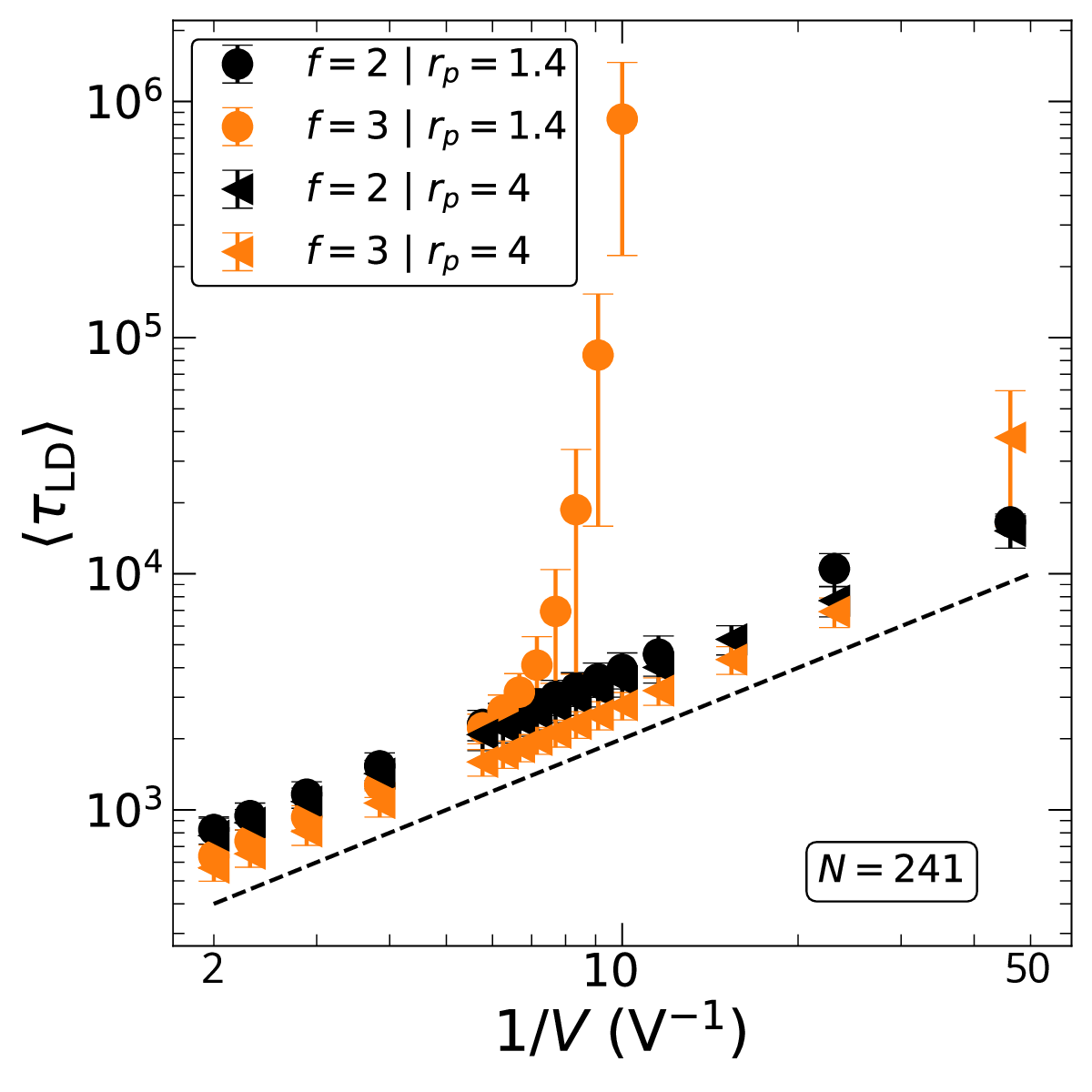}
        \caption{ }
        \label{fig:tau_LD_N241}
    \end{subfigure}
    \caption{Dependence of the mean translocation time, $\langle \tau_{\mathrm{LD}} \rangle$, on the applied transmembrane voltage, $V$, for different pore radii $r_p$, polymers with functionalities $f=2$ and $3$ and lengths (a) $N=151$ and (b) $N=241$. The dotted line depicts the scaling relation $\langle \tau_{\mathrm{LD}} \rangle \propto V^{-1}$, observed for $f=2$. The data for $f=3$ shows deviation from the scaling relation at lower $V$. }
    \label{fig:tau_LD_N151_N241}
\end{figure}

The mean translocation time $\langle \tau_\text{LD} \rangle$ for a linear polymer ($f=2$) of length $N=151$ is shown as a function of inverse of the transmembrane voltage, i.e. $1/V$, for three different pore radii in Figure \ref{fig:tau_LD_N151} (black symbols). The mean translocation time for a three-arm star polymer of the same molecular weight as the linear polymer is also included in these figures. For the linear polymer, a scaling of $\langle \tau_\text{LD} \rangle \sim 1/V$ is observed over a wide range of voltages between $\sim 22-500$ mV for each pore radius simulated. The same scaling relation is observed for a longer linear polymer of $N=241$, as seen from the black symbols in Figure \ref{fig:tau_LD_N241}. This scaling has been consistently reported for linear polymers in several experimental, \cite{Storm2005,Wanunu2008,Liu2012,Carson2014,Chen2024} simulation, \cite{Katkar2018b,Sarode2025} and theoretical studies. \cite{Muthukumar2003,Muthukumar2011PolymerTranslocation}. On the contrary, the mean translocation time of the three-arm star polymer exhibits the inverse scaling behavior only at high voltages, and deviates from it at lower voltages. For example, the scaling of $\langle \tau_\text{LD} \rangle \sim 1/V$ is observed only for $V\geq250$ mV for $N=151$ and $r_p=1.4$, while a pronounced deviation from this scaling behavior is already seen at $V\leq173$ mV. This deviation is observed for all combinations of $r_p$ and $N$ simulated. The value of the voltage below which a pronounced deviation from the $1/V$ scaling behavior is observed is found to be dependent on the nanopore radius $r_p$ and the molecular weight $N$ of the polymer. 

\subsection{Free energy: Metadynamics Simulations}
\label{sec:Free_energy}

\begin{figure}
    \centering
    \begin{subfigure}[b]{0.48\linewidth}
        \centering
        \includegraphics[width=\linewidth]{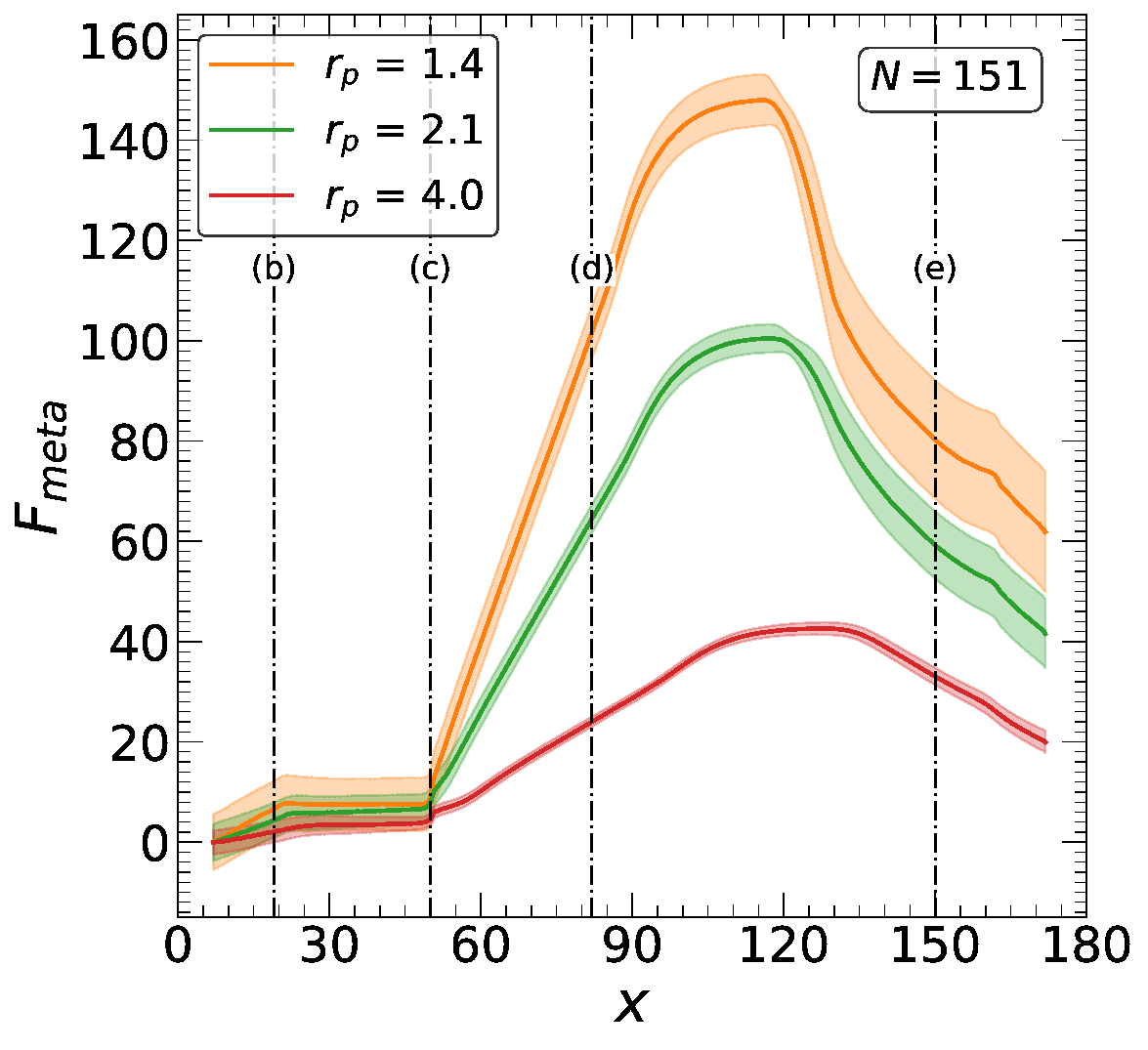}
        \caption{ }
        \label{fig:FE_f3_PL_151}
    \end{subfigure}
    \hfill
    \begin{subfigure}[b]{0.48\linewidth}
        \centering
        \includegraphics[width=\linewidth]{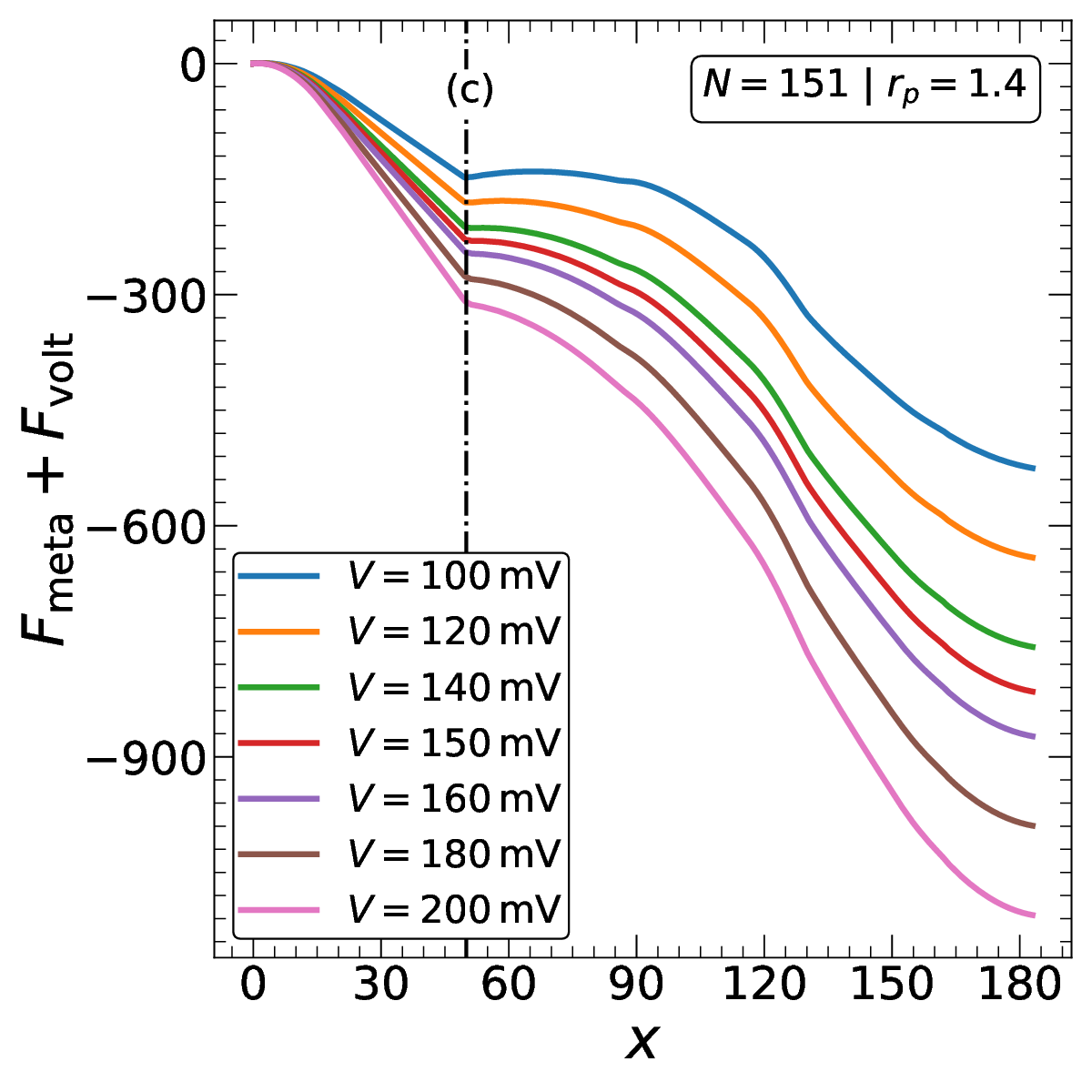}
        \caption{ }
        \label{fig:tilted_FE_N151}
    \end{subfigure}
    \caption{Free-energy landscapes, $F_\text{meta}$, for three-arm star polymers ($f=3$) translocating through nanopores of different radii, $r_p$, indicated in the legend, for a polymer length of $N=151$. The dash-dotted vertical lines mark the onset of the different translocation stages illustrated in Fig.~\ref{fig:schematic_trans_progress_curve}. (a) Free-energy landscapes, $F_\text{meta}$, at zero applied voltage ($V=0$) for $r_p=1.4$, 2.1, and 4.0. (b) Voltage-biased free-energy landscapes, $F_\text{meta}+F_\text{volt}$, for $r_p=1.4$ at different applied voltages, as indicated in the legend. The dash-dotted vertical line at the reaction coordinate $x=50$ marks the onset of branch-point transfer Stage of the translocation process.}
    \label{fig:FE_tilted_FE_N151}
\end{figure}

To investigate the effect of this distinct scaling behavior for $f=3$ relative to $f=2$, it is necessary to investigate the free energy landscape of the process. The free energy in absence of any voltage, $F_\text{meta}$ is shown in Figure \ref{fig:FE_f3_PL_151} for a three-arm star polymer of length $N=151$ translocating through nanopores of three different radii, i.e. $r_p=1.4$, $r_p=2.1$ and $r_p=4.0$. The vertical dotted lines show states that are illustrated in Figure \ref{fig:schematic_trans_progress_curve}. $F_\text{meta}$ increases nearly linearly during the pore filling stage, and remains nearly flat during the leading-arm threading stage. As the branch point enters through the nanopore and is transferred across the nanopore, $F_\text{meta}$ further increases, with a steeper slope compared to the pore filling stage. During the trailing-arm threading stage, $F_\text{meta}$ gradually increases, then remains constant, followed by a decrease. The pore emptying stage is characterized by a falling $F_\text{meta}$. Two prominent barriers can be identified in $F_\text{meta}$, one at the end of the pore filling stage, and the other during the trailing-arm threading stage. The first barrier is entropic in nature, and results from the loss of entropy of the beads inside the nanopore. Understandably, the height of this barrier decreases with increasing $r_p$. Such a decrease in the entropic barrier is also observed for a linear polymer, as shown in our earlier work \cite{Sarode2025}. The second barrier is significantly larger compared to the first barrier, with both, an entropic contribution and an additional contribution coming from the electrostatic repulsion between beads belonging to two arms that are present inside the nanopore. The height of the second barrier is found to decrease with increasing $r_p$ along with a slight shift towards higher $x$. $F_\text{meta}$ landscapes for intermediate pore radii are included in Figure SI-F4(a) of the supplementary document. The supplementary document Figure SI-F4(b) shows the free energy landscape for a longer polymer with $N=241$. The features observed in $F_\text{meta}$ for $N=151$ are also seen for $N=241$. The leading-arm threading stage and the trailing-arm threading stage are both extended due to the longer arms of the $N=241$ polymer. As expected, at constant $r_p$, the height of the first barrier, which arises from the loss of entropy of the beads inside the nanopore, is not found to be dependent on $N$. Similarly, the height of the second barrier remains nearly the same for the two values of $N$ at constant $r_p$.

$F_\text{meta}$ gets tilted upon adding the voltage contribution $F_\text{volt}$ using equation \ref{eq:Fvolt}. Figure \ref{fig:tilted_FE_N151} shows the voltage-dependent free energy $F=F_\text{meta} + F_\text{volt}$ for different applied voltages in the range from 100 mV to 200 mV. Note that for the polymer and pore, Langevin dynamics simulation results show a pronounced deviation from the scaling $\langle \tau_\text{LD} \rangle \sim 1/V$ (circles in Figure \ref{fig:tau_LD_N151}) for $V\leq 173$. The free energy at every $x$ decreases as the voltage increases. More importantly, the strengths of both the barriers decreases upon increasing the voltage. The first free-energy barrier corresponding to the pore-filling stage observed in Figure \ref{fig:FE_f3_PL_151} at V=0 significantly reduces and shifts towards the left upon increasing the voltage, as seen from Figure \ref{fig:FE_tilted_FE_N151}. For example, the barrier reduces and shifts from nearly $9$ at $x=16$ observed at $V=0$ to $0.7$ at $x=2.3$ for $V=100$ mV, and further to $0.4$ at $x=1.3$ for $V=180$ mV. The reduction in the barrier height and shift in its location is even more prominent in free energy at all higher voltages shown. Thus, the free energy is mostly downhill in the range of $x\in[0,50]$ (pore filling and leading-arm threading stages) at all these voltages. The branch point enters the nanopore at $x=50$, and transfers across the nanopore until $x=82$. At $V=0$, the entire branch-point transfer stage is uphill in free energy. For $V=100$ mV, the free energy in this range of $x$ is less steep, but still remains uphill. Thus, the height of the second barrier reduces upon increasing $V$ until 160 mV. At $V=180$ mV, the free energy becomes purely downhill in this range of $x$, and the second barrier disappears. Thus, at $V>180$ mV, the entire free energy is downhill, except for the tiny first barrier near $x=0$. 

Physically, at lower voltages, the resisting force resulting from the entropic penalty and the electrostatic repulsion between trailing arm beads confined inside the nanopore is dominant compared the driving force due to the applied voltage during the branch-point transfer stage. The driving force increases with increasing voltage, and dominates this resisting force beyond a threshold. For the given polymer and pore, the threshold appears to be between $160-180$ mV. Due to the presence of a large second barrier, the mean translocation time is expected to be significantly delayed for lower voltages, as seen in the next section.

\subsection{Translocation Kinetics: Fokker-Planck Theory}
\label{sec:translocatio_kinetics_FP}

The mean exit time from the right end, $\langle \tilde{\tau}_{\text{FP}} \rangle$ and its standard deviation are calculated from equations \ref{eq:dist_fpt}-\ref{eq:sd_fpt} for each voltage $V$. The system's diffusivity along the reaction coordinate $k_\text{FP}$ is used to fit the resulting unscaled mean exit time $\langle \tau_{\text{FP}} \rangle=\langle \tilde{\tau}_{\text{FP}} \rangle/k_\text{FP}$ to the mean translocation time data, $\langle \tau_{\text{LD}} \rangle$ at all voltages studied, by minimizing the error
\begin{align}
    \text{error}(k_\text{FP}) = \sum_{V}^{} \left( \frac{1}{k_\text{FP}}\langle \tilde{\tau}_\text{FP} \rangle - \langle \tau_{\text{LD}} \rangle \right)^2
    \label{eq:kfp_minimize}
\end{align}

\begin{figure}
    \centering
    \includegraphics[width=0.5\linewidth]{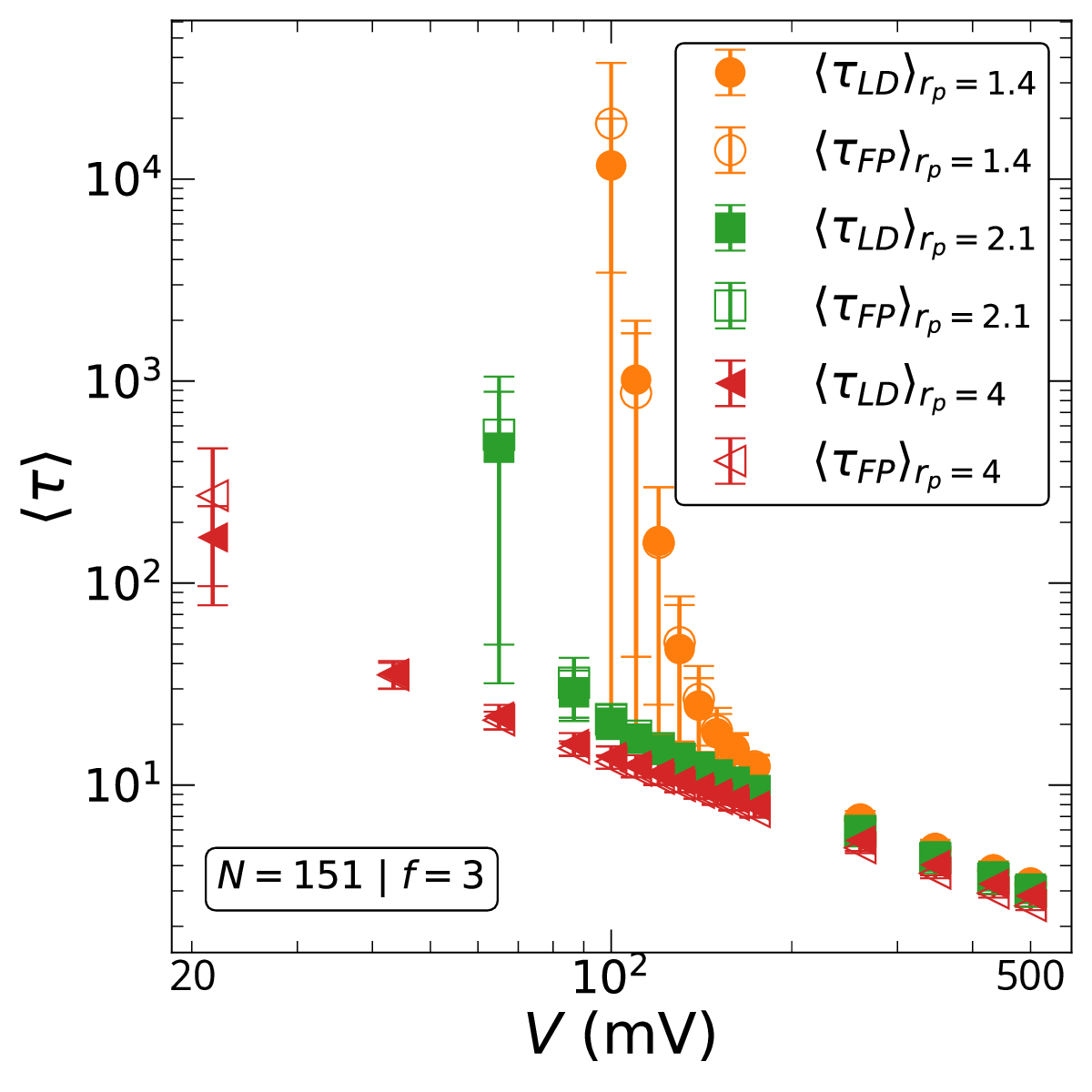}
    \caption{A comparison of mean translocation time $\langle \tau_\text{LD} \rangle$ (filled symbols) and mean exit time $\langle \tau_\text{FP} \rangle$ (open symbols) at different voltages $V$, for a polymer of length $N=151$ and varying pore radii $r_p=1.4, 2.1$ and $4$.}
    \label{fig:tau_LD_FP_N151}
\end{figure}

Figure \ref{fig:tau_LD_FP_N151} shows a remarkable agreement between the mean translocation time $\langle \tau_\text{LD} \rangle$ (filled circles) and the mean exit time $\langle \tau_\text{FP} \rangle$ (open circles) for the range of voltages studied $V$, for a polymer of length $N=151$ and for a pore radius of $r_p=1.4$, using a single value of $k_\text{FP} = 1.6804\times 10^{10}$ s$^{-1}$. The theory also predicts a scaling of $\langle \tau_\text{FP} \rangle \sim 1/V$ for voltages above 180 mV, and a deviation from this scaling at lower voltages. A similar agreement between $\langle \tau_\text{LD} \rangle$ and $\langle \tau_\text{FP} \rangle$ is also seen for $r_p=2.1$ and $4$. The respective $k_\text{FP}$ values are reported in Table \ref{tab:kFP}. The agreement between the model predictions for additional pore radii is shown in the supplementary document, Figures SI-F8 and SI-F9. The same agreement is also observed for $N=241$, as shown in the supplementary document Figure SI-F10.

Consider the second barrier in the free energy $F$ for $N=151$, $r_p=1.4$, and $V=100$ mV shown in Figure \ref{fig:tilted_FE_N151}. The solution $W(x,t)$ of the Fokker-Planck equation \ref{eq:non_dimen_FPE} is used in order to calculate the evolution of the flux exiting from the right end, $J(x=N+(f-1)M,t)$. The initial probability $W(x,t=0)$ is nearly a Dirac delta function at $x=2$, numerically approximated using a Gaussian distribution function. Upon integrating the Fokker-Planck equation, the distribution spreads due to diffusion, with its peak shifting due to the drift resulting from the local free energy gradient. The downhill free energy for $x<50$ results into a forward drift, while presence of the second barrier at $x>50$ results into a backward drift towards $x=50$. This results into accumulation of the probability at $x=50$ for a long time, before it escapes from either boundary. Such prolonged accumulation of the probability is not observed at voltages above a threshold. Instead, the initially narrow Gaussian distribution function keeps on spreading and drifting towards right with time. Evidently, we also observe that entry of the branch-point into the nanopore is relatively delayed for lower voltages in our Langevin dynamics simulation, as seen in supplementary video.

Because the features in $F_\text{meta}$ for the branch-point transfer stage, including the height of the second barrier is found to be nearly independent of $N$, the threshold voltage above which prolonged accumulation of the probability is not observed is also expected to be nearly independent of $N$ for the same $r_p$. Thus, the pronounced deviation from the $\langle \tau \rangle \sim 1/V$ behavior for $N=151$ in Figure \ref{fig:tau_LD_FP_N151} and $N=241$ in supplementary document Figure SI-F10 for a given $r_p$ is observed at the same voltage.

\begin{figure}
    \centering
    \includegraphics[width=0.5\linewidth]{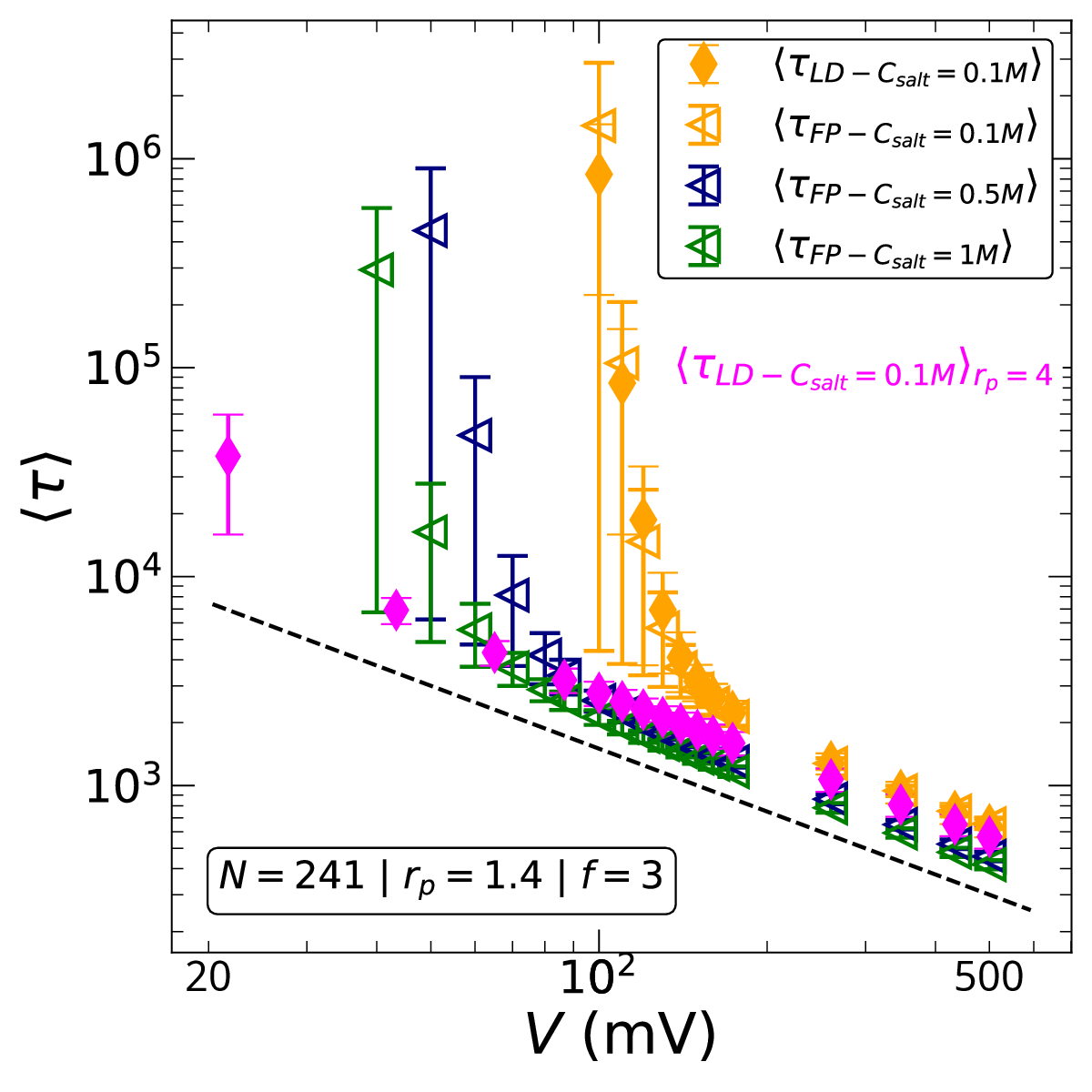}
    \caption{Mean translocation time $\langle \tau_\text{LD} \rangle$ (filled diamonds)and mean exit time $\langle \tau_\text{FP} \rangle$ (open left triangles) variation with the applied transmembrane voltage $V$ for the polymer of length $N=241$, pore radii $r_p=1.4$ and $4$ and monovalent salt concentration of $C_\text{salt} = 0.1, 0.5$ and $1.0$ M.}
    \label{fig:kinetics_comparision_rp1.4_PL_241_Csalt}
\end{figure}

The contribution of electrostatic repulsion between trailing arm beads during the branch-point transfer can be tweaked using salt concentration. We performed Langevin dynamics simulation for $N=241$ and $r_p=1.4$ at two additional salt concentrations, $C_\text{salt}  =0.5$ M and $C_\text{salt}  = 1$ M. The inverse Debye length used in $U_\text{DH}$ potential was modified to $\kappa=0.6886$ and $\kappa=0.9739$, respectively. Metadynamics simulation runs were also performed for these values of $\kappa$. The effect of salt concentration on $F_\text{meta}$ is shown in supplementary document Figure SI-F5. As expected, the second barrier decreases with increasing $C_\text{salt}$, due to a reduction in the electrostatic repulsion between trailing arm beads during the branch-point transfer stage. Figure \ref{fig:kinetics_comparision_rp1.4_PL_241_Csalt} shows the mean translocation time from Langevin dynamics simulation and the mean exit time predicted by the Fokker-Planck model. The corresponding values of the fitting parameter, $k_\text{FP}$ are included in Table \ref{tab:kFP}. The deviation from the $\langle \tau \rangle \sim 1/V$ scaling is observed at progressively lower voltages with increasing salt concentration. 

\begin{table}[ht]
\centering

\begin{tabular}{|c|c|c|c|}
\hline
\textbf{Polymer} &
\textbf{Pore Radius} &
\textbf{Salt Concentration} &
\textbf{$k_{\mathrm{FP}}$} \\
\textbf{Length, $N$} &
\textbf{$r_p$} &
\textbf{$C_{\mathrm{salt}}$ (M)} &
\textbf{($\times10^{10}\,\mathrm{s}^{-1}$)} \\
\hline

\multirow{7}{*}{151}
& 1.4 & 0.1 & 1.6804 \\ \cline{2-4}
& 1.7 & 0.1 & 1.7164 \\ \cline{2-4}
& 1.8 & 0.1 & 1.7312 \\ \cline{2-4}
& 2.1 & 0.1 & 1.7859 \\ \cline{2-4}
& 3.2 & 0.1 & 1.8502 \\ \cline{2-4}
& 3.8 & 0.1 & 1.8791 \\ \cline{2-4}
& 4.0 & 0.1 & 2.0531 \\
\hline

\multirow{4}{*}{241}
& \multirow{3}{*}{1.4}
& 0.1 & 1.3807 \\ \cline{3-4}
&      & 0.5 & 1.9234 \\ \cline{3-4}
&      & 1.0 & 2.0811 \\ \cline{2-4}
& 4.0  & 0.1 & 1.5271 \\
\hline
\end{tabular}
\caption{Values of the diffusivity, $k_{\mathrm{FP}}$, obtained from the Fokker--Planck model for different polymer lengths ($N$), nanopore radii ($r_p$), and salt concentrations ($C_{\mathrm{salt}}$).}
\label{tab:kFP}
\end{table}

\begin{figure}
    \centering
    \includegraphics[width=0.5\linewidth]{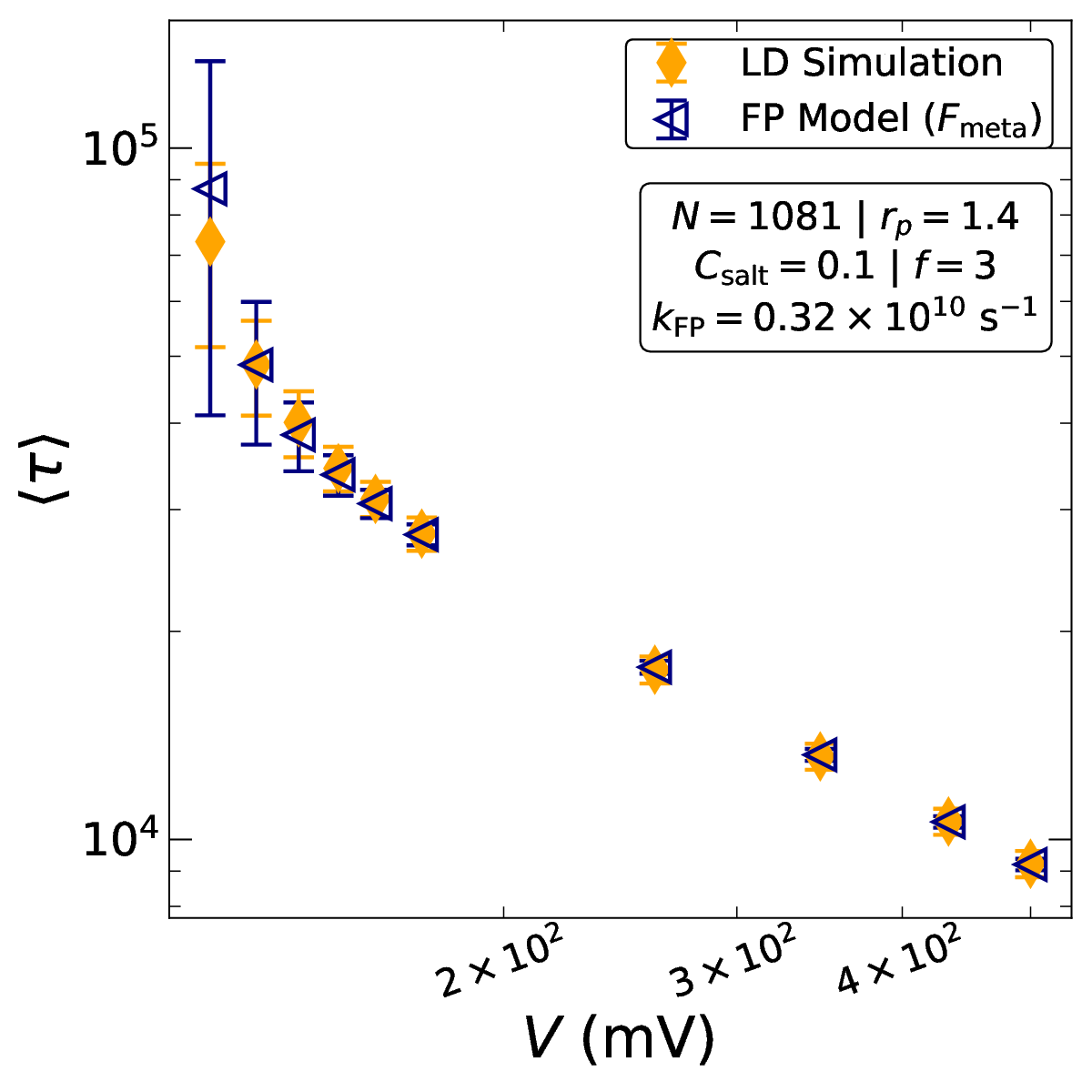}
    \caption{Comparison of mean translocation time $\langle \tau_\text{LD} \rangle$ (filled symbols) and the predicted mean exit time $\langle \tau_\text{FP} \rangle$ (open symbols) for a polymer of length $N=1081$, obtained by extending the free energy landscape of $N=241$ for pore radius of $r_p=1.4$ and salt concentration of $C_\text{salt}=0.1$. The calculated diffusivity $k_\text{FP} (= 0.32 \times 10^{10}$ s$^{-1})$ is shown.}
    \label{fig:tau_LD_FP_N1081}
\end{figure}

Comparing the estimated $F_\text{meta}$ for $N=151$ and $N=241$ at a given $r_p$ and $C_\text{salt}$, we find that the key differences are only in the length of the leading-arm threading stage and the trailing-arm threading stage, while overall features of the free energy are nearly the same. Further, the two lengths are related to $N$ through the arm length $\ell = (N-1)/3$. Using this intuition, we extend $F_\text{meta}$ of $N=241$ at $r_p=1.4$ and $C_\text{salt}=0.1$ M for higher molecular weight polymers. The extended free energy for $N=1081$ is shown in supplementary document Figure SI-F6. A comparison of the mean translocation time calculated from Langevin dynamics simulation and Fokker-Planck theory using the extended free energy is shown in Figure \ref{fig:tau_LD_FP_N1081}. The theoretical prediction using the extended free energy agrees reasonably well with the Langevin dynamics simulation data.

\section{Conclusion}

A central finding of this study is that for a three-arm star polymer with its arms longer than the nanopore length, the scaling of the mean translocation time  exhibits a pronounced deviation from the conventional relation $\langle \tau \rangle \propto V^{-1}$ below a threshold voltage. This scaling has been widely reported for linear polymers in experimental, theoretical, and simulation literature, and is observed in our own simulations. We find that the threshold voltage is nearly independent of the molecular weight of the three-arm star polymer, but varies with the nanopore radius and the salt concentration. We demonstrate that this anomalous scaling arises from a unique second free-energy barrier that is observed for a three-arm star polymer. This barrier is associated with confining charged trailing arms inside the narrow nanopores, and is a result of entropic and electrostatic contributions. At low applied voltages, this additional free-energy barrier slows-down the translocation kinetics, whereas at increasingly high voltages, the conventional scaling is gradually recovered as the external driving force overcomes the barrier. Lastly, a simple extension of the free energy landscape is shown to make a reasonably accurate prediction of translocation kinetics for higher molecular weight polymers.

\begin{acknowledgments}
This work was supported in part by the Initiation Grant at Indian Institute of Technology Kanpur and in part by Science and Engineering Research Board (SERB) through the Start-up Research Grant Number SRG/2020/002186. The resources provided by PARAM Sanganak under the National Supercomputing Mission, Government of India at the Indian Institute of Technology, Kanpur are gratefully acknowledged.

\end{acknowledgments}

\section*{Supplementary Material}

This article contains a supplementary document and a supplementary video. The supplementary document contains details of metadynamics simulation, numerical methods and translocation kinetics for additional molecular weight of polymer, size of nanopore and salt concentration.

\bibliography{bibexport2}

\end{document}